\documentclass{aa}  

\usepackage{graphicx}
\usepackage{txfonts}
\begin{document}

   \title{New insights into the proton precipitation sites in solar flares}
   %\title{Proton precipitation sites in solar flares through $\gamma$-ray and white-light observations}
   %\titlerunning{aa}

   \author{Andrea Francesco Battaglia \inst{1}
          \and
          Säm Krucker \inst{2,3,4}
          }

   \institute{
        Istituto ricerche solari Aldo e Cele Daccò (IRSOL), Faculty of informatics, Università della Svizzera italiana, Locarno, Switzerland \\
        \email{andrea.francesco.battaglia@irsol.usi.ch}
        \and
        University of Applied Sciences and Arts Northwestern Switzerland (FHNW), 5210 Windisch, Switzerland
        \and
        Space Sciences Laboratory, University of California, 94720 Berkeley, USA
        \and
        ISEE, Nagoya University, Japan}

   \date{Received 25 November 2024 / Accepted 16 December 2024}

% \abstract{}{}{}{}{} 
% 5 {} token are mandatory
\abstract
   {}
  % aims heading (mandatory)
   {We revisit the Reuven Ramaty High Energy Solar Spectrocopy Imager (RHESSI) $\gamma$-ray observations of the extraordinary GOES X25 flare SOL2003-10-28T11:10 to investigate previously reported conclusions that flare-accelerated electrons and protons precipitate along spatially separated flare loops.}
  % methods heading (mandatory)
   {In contrast to previous works that reconstructed 2.223 MeV images over extended time periods ($\sim$20 minutes), we selected shorter integration times on the order of 2 to 3 minutes. Using simulations of the 2.223 MeV profile in combination with observations of the prompt $\gamma$-ray lines from the INTEGRAL mission, we obtained two separated integration time ranges representing the peak of the flare and the start of the decay, respectively. The resulting $\gamma$-ray images were then compared to GONG white-light (WL) observations to identify where along the flaring ribbons electron and proton precipitation occurs.}
  % results heading (mandatory)
   {We point out that previously reported results comparing RHESSI hard X-ray (HXR) and $\gamma$-ray images only hold if the relative time evolution in the two energy ranges is the same. 
   As the decay time for the October 28, 2003, flare is different at the two energy ranges considered (200-300 keV and around 2.223 MeV), the previously published conclusion that electrons and protons precipitate at different locations is an overstatement. Using shorter integration times reveals that the $\gamma$-ray and HXR sources spatially coincide with the WL flare ribbons.}
  % conclusions heading (optional), leave it empty if necessary 
   {Our key conclusion is that electron and proton precipitation sites coincide with the flare ribbons, suggesting that the electron and proton precipitation sites are the same, at least within RHESSI's imaging capabilities. This result solves the 20-year-long mystery surrounding the different previously reported electron and proton precipitation sites.}


   \maketitle
%
%-------------------------------------------------------------------

\section{Introduction}

Solar flares, the most powerful release of energy in the entire Solar System, emit radiation across the entire electromagnetic spectrum, from radio, to visible, up to hard X-rays (HXRs) and $\gamma$-rays \citep{2011SSRv..159...19F}. Through the release of “free” magnetic energy, high-energy electrons, protons, and ions are accelerated in the low corona, and can either travel along magnetic field lines toward the solar surface or escape into interplanetary space. The downward flare-accelerated particles, upon interaction with the much denser chromosphere and photosphere, lose most of their energy by collision (i.e., heating) or by radiation. In the HXR and $\gamma$-ray ranges, part of this radiation is generated by semi-relativistic and relativistic ($\gtrsim 300 - 400 \, \mathrm{keV}$) electrons through continuum bremsstrahlung emission. Additionally, protons and ions produce spectral lines in the $\gamma$-ray range through nuclear reactions and particle decay processes. For instance, upon collision with ambient ions, different nuclear de-excitation lines are produced from $\sim 0.8 \, \mathrm{MeV}$ to $\sim 10 \, \mathrm{MeV}$, such as $^{12}$\ion{C}{} at $4.438 \, \mathrm{MeV}$ and $^{16}$\ion{O}{} at $6.129 \, \mathrm{MeV}$. These broad de-excitation lines merge into a quasi-continuum dominating the electron bremsstrahlung continuum \citep{2011SSRv..159..167V}. Because the de-excitation after collision has a timescale of $10^{-12} \, \mathrm{s}$ \citep[][]{2011SSRv..159..167V}, these lines are known as “prompt” lines, as there is no a measurable delay and they are thus emitted almost instantaneously. The neutron-capture line, emitted at 2.223 MeV upon the capture of a neutron by ambient hydrogen (forming deuterium), has a delay between the generation of the neutron and its capture \citep[e.g.,][]{2003ApJ...595L..93M,2011SSRv..159..167V}. The reason for this delay is the thermalization of the neutrons, because the scattering cross section of a flare-produced neutron is larger than that of a neutron-capture \citep{1987SoPh..107..351H}. Therefore, a flare-produced neutron first undergoes a number of collisions (i.e., thermalization) before being captured and consequently emitting the 2.223 MeV line. The timescale of the generation of the neutron-capture line is about 100 s \citep{2007ApJS..168..167M}. However, shorter timescales have been found, such as in \citet{2019ApJ...877..145L}, which found a delay of about $10 \, \mathrm{s}$. The capture probability is highest for neutrons produced in the photosphere, where they are least able to escape or decay before they are thermalized and captured \citep{1987SoPh..107..351H,1989SoPh..121..323R}.

The history of HXR imaging observations of flares is marked by a successful heritage of space-based imaging telescopes. Some examples, among others, include the Hard X-ray Telescope \citep[HXT;][]{1992PASJ...44L..45K} aboard the Yohkoh mission, the Reuven Ramaty High-Energy Solar Spectroscopic Imager \citep[RHESSI;][]{2002SoPh..210....3L}, and the currently operational Spectrometer/Telescope for Imaging X-rays \cite[STIX;][]{2020A&A...642A..15K}, as well as the Hard X-ray Imager \citep[HXI;][]{2019RAA....19..160Z} aboard the Advanced Space-based Solar Observatory \citep[ASO-S][]{2019RAA....19..156G}. Imaging $\gamma$-rays in flares, instead, is notoriously much more challenging. These challenges arise from the combination of technological difficulties in detecting and imaging $\gamma$-rays, which require significantly more mass in the space instrument than HXRs, and counting statistics issues due to the steep decrease in the spectrum from HXRs to $\gamma$-rays. This decrease amounts to approximately two to three orders of magnitude in flux for every tenfold increase in energy. Consequently, only flares that are highly efficient in accelerating high-energy protons and ions can produce detectable $\gamma$-rays, typically corresponding to the most energetic events \citep[e.g.,][]{2009ApJ...698L.152S}. Currently, no operational solar $\gamma$-ray imaging telescope exists for routine flare observations. Nevertheless, efforts in developing future instruments are being undertaken in order to address this gap in our observational capabilities \citep[e.g.,][]{2023Aeros..10..985R}.

Despite the challenges surrounding the detection of $\gamma$-rays, protons and ions constitute key components in flares. For instance, based on the analysis of a few dozen $\gamma$-ray producing flares, it was found that the energy contained in $>1$ MeV ions lies in the range of $10^{29-33}$ ergs \citep[e.g.,][]{1995ApJ...455L.193R,1997JGR...10214631M,1997ApJ...479..458R,1997ApJ...490..883M,2011SSRv..159..167V}. This shows that, beyond the electrons probed in the HXR range, diagnosing ions in the $\gamma$-ray range in flares is of the utmost importance, as the energy content of flare-accelerated electrons and ions are comparable \citep{2012ApJ...759...71E}.

The first-ever and, to date, the only $\gamma$-ray images of solar flares were obtained by RHESSI. Even though RHESSI observed countless energetic flares, imaging of $\gamma$-ray emission was only possible for about half a dozen events \citep{2003ApJ...595L..69L,2006ApJ...644L..93H,2008ApJ...678L..63K,2011SSRv..159..167V}. Nevertheless, many important results have been obtained and we refer to \citet{2011SSRv..159..167V} for a comprehensive review. The first-ever $\gamma$-ray image of a flare, of the X4 GOES class, occurred on July 23, 2002, and is reported in \citet{2003ApJ...595L..69L}. In this paper, the authors showed that the centroid of the ion-produced 2.223 MeV neutron-capture line emission is located approximately $20\pm6$ arcsec away from the electron bremsstrahlung source locations, suggesting that the acceleration and/or propagation of the ions is different from that of the electrons. This result was further corroborated by \citet{2006ApJ...644L..93H}, who analyzed the source location of the 2.223 MeV neutron-capture $\gamma$-ray line in three large RHESSI flares of 2003 (October 28, October 29, and November 2), and reported a systematic displacement from the corresponding 0.2–0.3 MeV electron-bremsstrahlung emission footpoints of 10 to 25 arcsec. According to the standard flare model, this spatial separation remains puzzling, as both electrons and protons are assumed to be accelerated in the same region. Therefore, the question of why they propagate along different magnetic loops (and consequently precipitate at different locations) is still unresolved and could not be investigated with new observations. In this paper, we revisit the X25 GOES class flare that occurred on October 28, 2003,\footnote{The October 28, 2003, event was previously categorized as of the X17 GOES class. However, considering the recently established correction factor of previous GOES missions \citep{2024SoPh..299...39H}, the class of this event is estimated to be X25.} by incorporating additional information obtained from visible light observations.

Visible light observations have become crucial in studying the precipitation sites of flare-accelerated electrons, as over the past decade significant progress has been made in this context.
Visible light enhancements associated with flares, also referred to as white-light (WL) emission, are observed as either continuum enhancements \citep[e.g.,][]{1992PASJ...44L..77H,2006SoPh..234...79H,2011SoPh..269..269M,2012ApJ...753L..26M,2015ApJ...802...19K,2016ApJ...816...88K,2016ApJ...816....6K,2018A&A...620A.183J} and/or strong emission from photospheric spectral lines \citep[e.g.,][]{2004SoPh..220...81A,2017ApJ...838...32Y}\footnote{In this paper, we use the term WL in order to refer to generic photospheric emission enhancement, without differentiating between line or continuum enhancement. Using this term, we want to emphasize the impact of flare-accelerated electrons or ions to the lower atmosphere.}. 
By analyzing a flare close to the limb, \citet{2012ApJ...753L..26M} found that the WL and HXR centroids closely match, which suggests that the observed WL emission mechanism is directly linked to the energy deposition by flare-accelerated electrons. A similar result was reported by \citet{2015ApJ...802...19K}, who analyzed three limb events. In addition, \citet{2016ApJ...816....6K} conducted a statistical analysis revealing that HXRs and WL emissions are correlated not only spatially but also in terms of peak time, intensity, and energy deposition, which highlights the crucial role of flare-accelerated electrons in WL formation. Therefore, WL observations in flares provide valuable insights into electron precipitation sites. To the best of the authors' knowledge, no observational studies have yet investigated if there is any relationship between the source location of $\gamma$-rays (i.e., the proton precipitation sites) and WL flare ribbons.

In this paper, we revisit the extraordinary X25 GOES class flare that occurred on October 28, 2003, which produced the most intense $\gamma$-ray signal ever imaged in solar flares. This event has been extensively investigated in numerous studies \citep[e.g.,][]{2005A&A...434.1183B,2006ApJ...644L..93H,2006A&A...445..725K,2006SoPh..236..325S,2007ApJ...664..573Z,2008ApJ...678L..63K,2008ApJ...678..509T,2024A&A...686A.195K}. Our objective is to analyze the $\gamma$-ray source locations of the 2.223 MeV neutron-capture line and contextualize it with WL emissions. This approach allows us to examine, for the first time, the correlation between WL emission and the proton precipitation sites. In Sect.~\ref{sec:data-analysis}, the data analysis and an overview of the event are presented. In Sect.~\ref{sec:results}, we report results and discussions. Finally, our conclusions are drawn in Sect.~\ref{sec:conslusions}. 

%--------------------------------------------------------------------

\section{Selected event and data analysis \label{sec:data-analysis}}

%----------

    \subsection{Event overview}

    \begin{figure}
        \centering
        \includegraphics[width=0.99\linewidth]{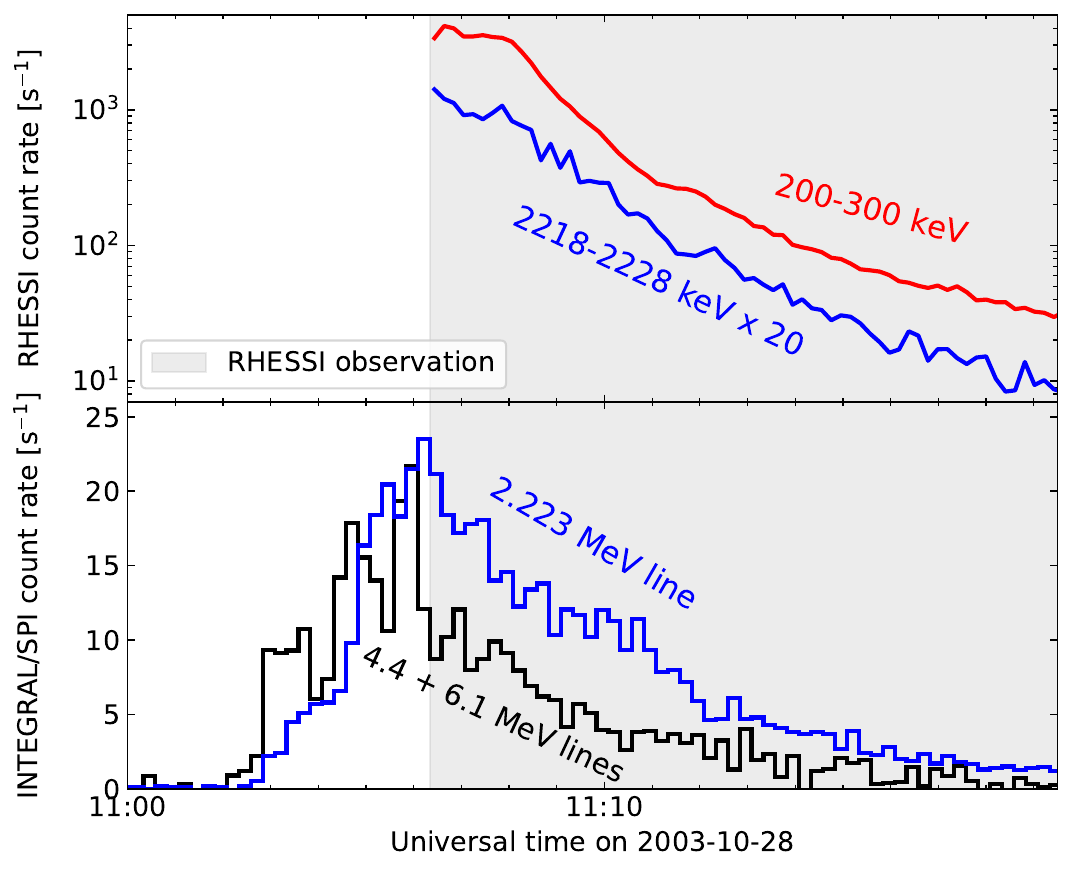}
        \caption{Time histories of the October 28, 2003, flare from RHESSI (\emph{top}) and INTEGRAL/SPI (\emph{bottom}). The gray area highlights the RHESSI coverage of the flare, which started right after the end of the peak of the impulsive phase at 11:06:20.}
        \label{fig:time-profiles}
    \end{figure}

    \begin{figure*}
        \centering
        \includegraphics[width=\linewidth]{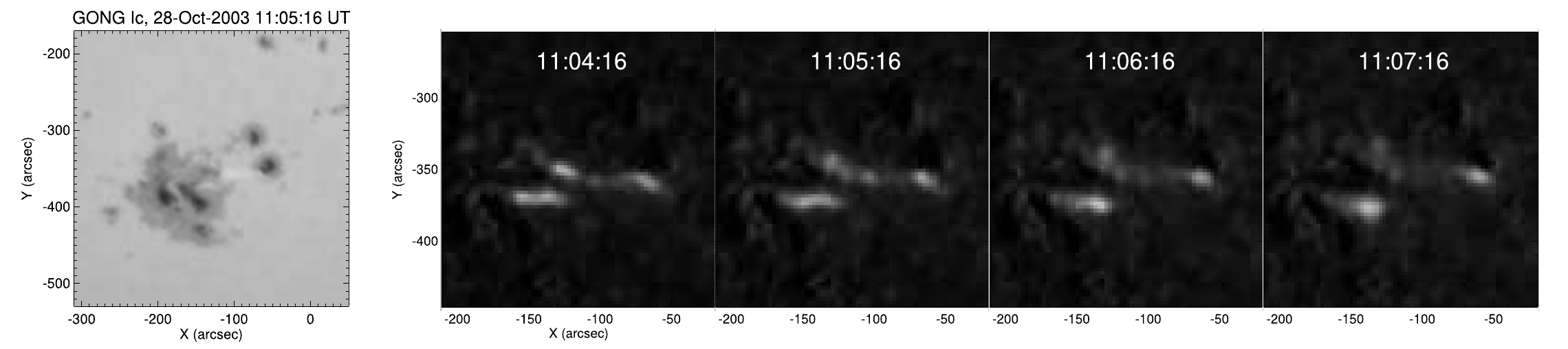}
        \caption{Overview of the GONG intensity observations of the October 28, 2003, flare. The leftmost panel shows the continuum intensity image of the region where the flare happened. From the second to the fifth panel, we show the time evolution (around the peak of the impulsive phase at 1 minute time resolution) of the pre-flare subtracted continuum intensity images. From the pre-flare subtracted images, we can clearly see the flare ribbons. The percent level of the maximum increase in WL emission due to the flare is 35\% of the pre-flare intensity.}
        \label{fig:GONG-figure}
    \end{figure*}

    To investigate the relationship between the $\gamma$-ray source locations of the 2.223 MeV neutron-capture line and WL emission, we focused on the X25 GOES class flare of October 28, 2003. This event, originating from AR 10486 near the disk center, produced the most intense $\gamma$-ray signal ever imaged in a solar flare. This flare has been observed by RHESSI. Although the spacecraft passage through the South Atlantic Anomaly prevented the rising and peak of the impulsive phase from being recorded, the exceptional magnitude of the flare allowed for significant HXR and $\gamma$-ray flux observations during the decaying impulsive phase. RHESSI fully captured this latter phase, as is illustrated in Fig.~\ref{fig:time-profiles}.
    
    The spectrometer instrument \citep[SPI;][]{2003A&A...411L..63V} aboard the International Gamma Ray Astrophysics Laboratory \citep[INTEGRAL;][]{2003A&A...411L...1W} provided comprehensive time profile observations for the entire flare event. In this study, we use SPI time profiles as a reference for the impulsive phase. The Global Oscillation Network Group telescope \citep[GONG;][]{1996Sci...272.1284H} in Cerro Tololo, Chile, also captured the impulsive phase. GONG observations are crucial for this study, as they provide information on the WL emission during both impulsive and decay phases, and thus reveal the electron precipitation sites. Fig.~\ref{fig:GONG-figure} presents an overview of the GONG continuum intensity observations. The pre-flare subtracted images clearly show the evolution of flare ribbons.

%----------

    \subsection{Data analysis}

    RHESSI data were acquired using the RHESSI software within IDL SolarSoftware (SSWIDL). Image reconstruction was performed using the CLEAN algorithm \citep{1974A&AS...15..417H} with natural weighting. For HXR images (200-300 keV), detectors 5 through 9 were used, while $\gamma$-ray images (2218-2228 keV) employed only detectors 6 and 9, which are associated with the thickest grids. Both HXR and $\gamma$-ray image reconstruction was done by considering counts from the rear detector segment only.

    The INTEGRAL/SPI data were taken from \citet{2006A&A...445..725K}.
    The GONG continuum intensity observations have been downloaded from the NSO GONG data archive\footnote{\url{https://gong2.nso.edu/archive/patch.pl?this_program=run_quick_day3&menutype=a&calendar=}}. These data product are available at one minute resolution. To obtain pre-flare subtracted continuum intensity images, we subtracted the observations taken one minute prior to flare onset from all subsequent observations.
    The Transition Region and Coronal Explorer \citep[TRACE;][]{1999SoPh..187..229H} 195 \AA{} extreme ultraviolet (EUV) observations were downloaded from the virtual solar observatory portal\footnote{\url{https://sdac.virtualsolar.org/cgi/search}}. 

%--------------------------------------------------------------------

\section{Results and dicussions \label{sec:results}}

%----------

    \subsection{White-light and extreme ultraviolet observations}

\begin{figure*}[!]
        \centering
        \includegraphics[width=0.8\linewidth]{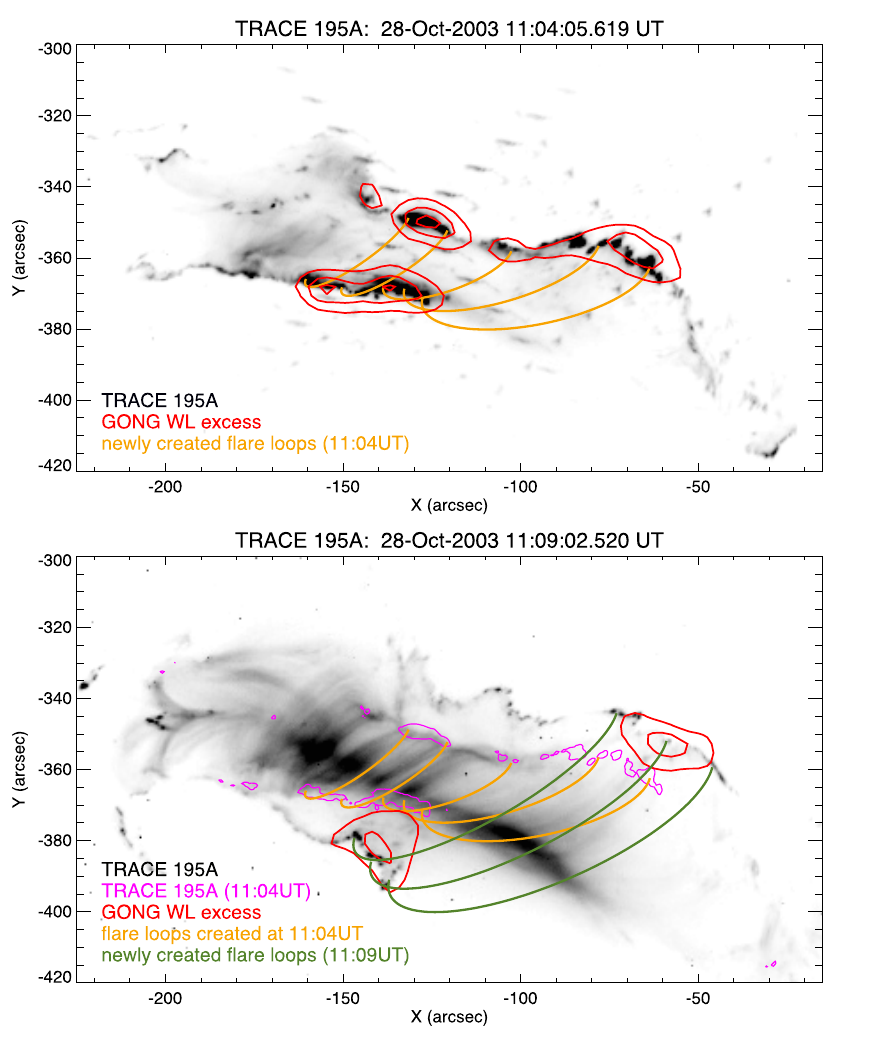}
        \caption{EUV and WL images of the flare at two different time instances. In both panels, the TRACE 195 \AA{} image is plotted as the background, with red contours overlaid to indicate the flare ribbons derived from pre-flare subtracted GONG intensity images (see Fig. 2). The two different times correspond to a time near the peak of the impulsive phase (\emph{top}) and a time during the decay phase (\emph{bottom}). To represent the flare loops connected to the ribbons, we draw as a proxy semicircles perpendicular to the solar surface, where yellow semicircles denote flare loops formed during the impulsive phase, while the green ones represent flare loops created during the decay phase. The selection of the footpoints for the proxy loops was driven to have several loops along the ribbons connecting the WL sources without much twist in them. The selection is somewhat arbitrary and intended as a visual guide.}
        \label{fig:loops-image}
    \end{figure*}

    The aim of this subsection is to compare EUV and WL images at two distinct times, as is illustrated in Fig.~\ref{fig:loops-image}. The upper panel displays EUV and WL images near the peak of the impulsive phase (11:04:05 UT). Notably, the bright EUV ribbons align well with the GONG WL contours. This combined EUV and WL emission clearly indicates the electron precipitation site, where energy is deposited into the lower atmosphere, chromosphere, and/or photosphere. To aid visualization, semicircles perpendicular to the solar surface have been drawn connecting the ribbons, highlighting the newly created flare loops.
    
    A similar pattern is observed at the later time (11:09:02 UT). The energy release and particle acceleration in this flare is long-lasting, which creates successively larger loops, as is illustrated in the bottom panel of Fig.~\ref{fig:loops-image}, as well as new ribbons at different locations.  
    These ribbons are associated with newly reconnected field lines (depicted as green loops), which align well with both EUV and WL emissions. Furthermore, the loops formed near the impulsive phase peak (shown in orange) now exhibit coronal emission in EUV, which have cooled down to become visible in the TRACE 194 \AA{} passband.  

    The EUV and WL observations of this event showcase a nice example that aligns with the standard flare reconnection model. In this model, magnetic field lines initially interact with field lines at a specific altitude, followed by subsequent reconnection at higher altitudes \citep[e.g.,][]{2009SoPh..255..107Q}. In addition, Fig.~\ref{fig:loops-image} illustrates how GONG WL observations effectively agree with the electron precipitation site, corroborating what is reported in previous studies \citep[e.g.,][]{2015ApJ...802...19K,2016ApJ...816....6K}.

%----------

    \subsection{Hard X-ray and $\gamma$-ray observations}

    In light of the recent advances in understanding WL generation and its correlation with HXRs, the comparison between WL and $\gamma$-ray emissions is central to this paper. By revisiting the $\gamma$-ray imaging of the October 28, 2003, event, we gained important insights into two distinct concepts: the challenge of comparing images at different energies reconstructed over extended time periods, and the varying contributions of different flare phases to the generation of the neutron-capture line. Along with the results obtained by revisiting the $\gamma$-ray imaging of this event, these two concepts are discussed in the following.

%----------

        \subsubsection{The extended time integration of previous studies}
        
        As a first step, we checked that the current version of the RHESSI software produces consistent images for RHESSI rear-segment images, as was reported in \citet{2006ApJ...644L..93H}. The RHESSI imaging software had improved over the years, with the main change being the introduction of visibilities \citep{2007ApJ...665..846P}. Due to these improvements, the reconstructed images are not exactly the same as reported in \citet{2006ApJ...644L..93H}, but the centroid location of the $\gamma$-ray sources are the same within the errors bars obtained from forward fitting. With the current software, by repeating the same analysis as was done in \citet{2006ApJ...644L..93H}, the resulting center positions for the two $\gamma$-ray sources reported are: $(x_{\mathrm{E}},y_{\mathrm{E}})=(-141^{\prime\prime},-388^{\prime\prime})$ for the eastern source and $(x_{\mathrm{W}},y_{\mathrm{W}})=(-75^{\prime\prime},-357^{\prime\prime})$ for the western source. These positions align closely with the centroid positions (one sigma uncertainty of $\pm 5^{\prime\prime}$) reported by \citet{2006ApJ...644L..93H}: $(x_{\mathrm{E}}^{\mathrm{H}},y_{\mathrm{E}}^{\mathrm{H}})=(-145^{\prime\prime},-380^{\prime\prime})$ for the eastern source and $(x_{\mathrm{W}}^{\mathrm{H}},y_{\mathrm{W}}^{\mathrm{H}})=(-74^{\prime\prime},-354^{\prime\prime})$ for the western source. The differences in coordinate values are up to $8^{\prime\prime}$, within the error range.
        
        The main difference of our analysis compared to previously published RHESSI $\gamma$-ray imaging results is in the selected time intervals and the new consideration of WL images. To enhance counting statistics as much as possible, the papers published in the past reconstructed images averaged over the entire duration of the flare. Comparing reconstructed RHESSI images of such long-time integrations is not as straightforward as it seems. The limited dynamic range and resolution of these reconstructions can be misleading if the relative time evolution in HXRs and $\gamma$-rays is different. To illustrate this, we assume a simple source geometry of a single, common source for electron and proton precipitation on a flare ribbon that moves from A to B during the integration time. Integrating over a long duration, the derived RHESSI source location is the flux-weighted average along the line between A to B. If the $\gamma$-rays decay slower than the HXRs, the gamma-ray source centroid is closer to B than for the HXR centroid. Hence, it appears that electrons and protons are precipitating on different locations, but this could be solely because the time profiles decay differently. Hence, the previously published results that electron and protons precipitate at different locations only holds if the electron and proton precipitation rates have the same time evolution. As both HXRs and $\gamma$-rays are used for imaging reconstruction, and their profiles are obviously different (refer to Fig.~\ref{fig:time-profiles}), diagnostics of source locations is, particularly for long integration times, ambiguous.

%----------

        \subsubsection{Different contributions to the neutron-capture line}

        \begin{figure}
            \centering
            \includegraphics[trim={20 250 20 250}, angle=-90,origin=c,width=1.\linewidth]{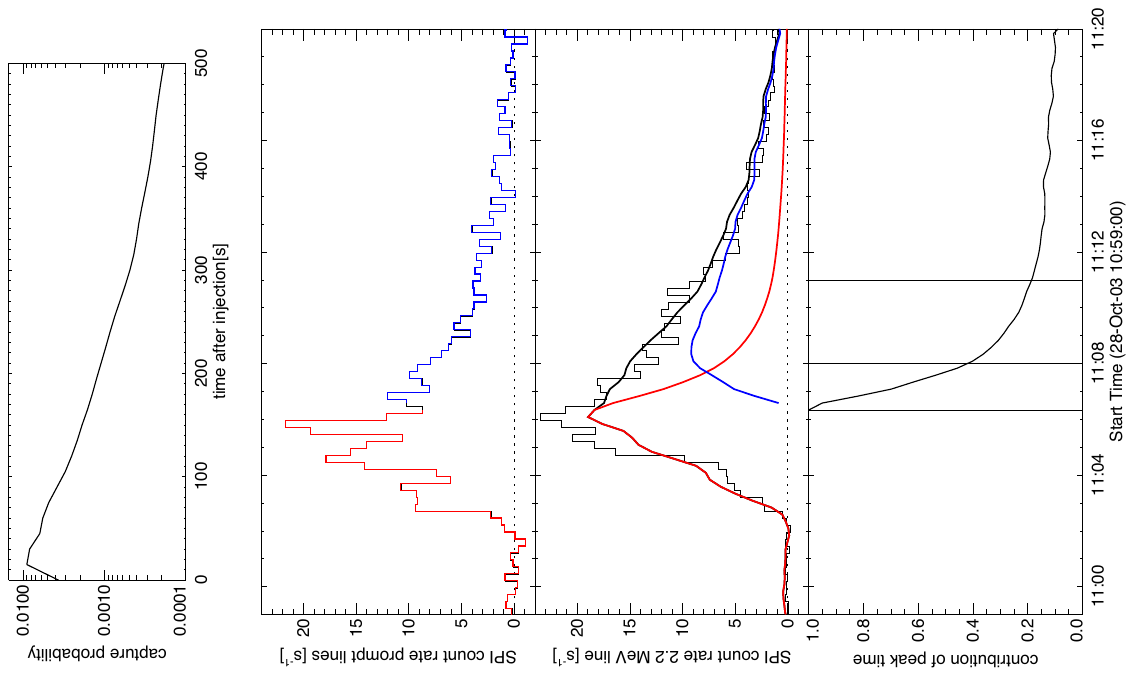}
            \caption{Relative timing of proton precipitation and creation of the 2.223 MeV neutron-capture line emission. (\emph{Top panel}) Time profile of the 2.223 MeV line emission, assuming a $\delta$-function injection of protons at time zero \citep[for details, see][]{2003ApJ...595L..93M}. (\emph{Second panel}) Time profiles of the INTEGRAL/SPI prompt lines, which is the same as the black curve in Fig.~\ref{fig:time-profiles}. Red indicates periods when RHESSI was not observing (peak of the impulsive phase), while blue highlights RHESSI observation times (decaying impulsive phase). (\emph{Third panel}) Time profiles of the INTEGRAL/SPI 2.223 MeV neutron-capture line (black histogram). Red indicates the contribution to the 2.223 MeV neutron-capture line of the times prior to RHESSI observations, and hence the peak of the impulsive phase, while blue indicates the contribution from the decaying impulsive phase. (\emph{Bottom panel}) Relative contribution to the 2.223 MeV neutron-capture line from the peak of the impulsive phase as a function of time, calculated from the ratio of the black curve in the middle panel divided by the red one. The vertical lines give the edges of the time intervals used for the images presented in Fig.~\ref{fig:rhessi-images}. }
            \label{fig:convolution-time-profiles}
        \end{figure}

        \begin{figure*}
            \centering
            \includegraphics[width=0.99\linewidth]{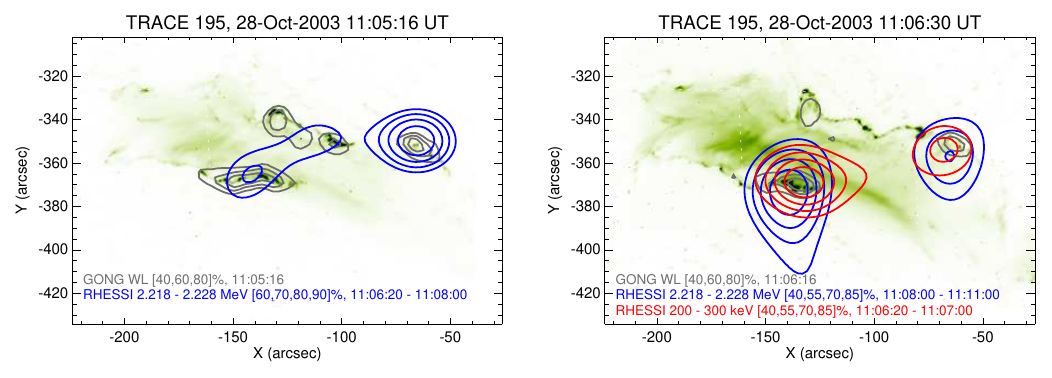}
            \caption{Multiwavelength imaging overview of the October 23, 2003, event at two different times. In both panels, the TRACE 195 \AA{} image is plotted as a green background, with gray contours representing flare ribbons derived from pre-flare subtracted GONG intensity images. Red contours indicate RHESSI HXR images, while blue contours show RHESSI $\gamma$-ray images. The left panel depicts a time near the impulsive phase peak, and the right panel illustrates a time during the decaying impulsive phase. The time integration intervals for the RHESSI $\gamma$-ray images have been adjusted based on the contribution of the different flare phases to the neutron-capture line (see Sect.~\ref{sec:results} for details). 
            }
            \label{fig:rhessi-images}
        \end{figure*}

        In addition to the challenge of comparing sources at different energies reconstructed over extended time periods, there is the intrinsic complication that the neutron-capture line at 2.223 MeV is not emitted instantaneously. The neutron-capture line emission is not instantaneous due to the thermalization process preceding capture, resulting in a time evolution with a relatively long decay. Consequently, careful evaluation of the contributions from various flare phases is essential during the reconstruction of $\gamma$-ray neutron-capture line images. Therefore, in our analysis, we propose a new approach that uses three main improvements:
        \begin{itemize}
            \item Much shorter integration times are used;
            \item Selected time intervals derived from observations of prompt $\gamma$-ray lines in combination with simulations of neutron-capture process \citep{2003ApJ...595L..93M};
            \item Comparison of the $\gamma$-ray images to WL observations by GONG at the time of the peak of the prompt $\gamma$-ray emission. 
        \end{itemize}
        
        \citet{2003ApJ...595L..93M} showed simulations of the actual delay of the neutron-capture line at 2.223 MeV produced by a $\delta$-function injections of protons. The obtained time profiles depended on several parameters such as the spectral index of the accelerated protons, the level of pitch angle scattering, the relative $^3$He abundances, and the flare location relative to the observer. The top panel of Fig.~\ref{fig:convolution-time-profiles} shows a typical profile taken from \citet{2003ApJ...595L..93M}. To qualitatively understand how the 2.223 MeV profile is produced, we used the observed prompt line profile from INTEGRAL/SPI as a measure of when protons enter the chromosphere or photosphere and convolved it with the typical profile from \citet{2003ApJ...595L..93M}. The resulting profile is shown in the second panel of Fig.~\ref{fig:convolution-time-profiles} in black compared with the observed 2.223 MeV signal from INTEGRAL/SPI. The two profiles show a similar behavior in time, confirming that we have a qualitative agreement. By varying the parameters of the simulation, a potentially better fit could be found. For example, selecting a simulation from \citet{2003ApJ...595L..93M} with a sharper, early peak will likely give a better agreement around peak time where the currently used profile smears out individual peaks. However, such an analysis is not the focus of our paper. Here, we have only used the derived profiles to check how well the 100 s approximation of the generation of the 2.223 MeV signal holds for this flare. The black curve is separated into two components: the blue curve is the part of the 2.223 MeV signal that is produced by protons that precipitate during the part of the decay time when RHESSI was observing. The red curve is the signal from protons that precipitate at times when RHESSI was not observing. Hence, 100 s after the start of the RHESSI observations, the 2.223 MeV signal contains about 20\% of the signal from the peak and 80\% from the decay. 
        This shows that the 100 s delay is a good approximation, considering the limited dynamic range of RHESSI imaging in the $\gamma$-ray range. Nevertheless, it is also clear that this approximation does not necessarily hold for all flares. For a flare with a with a weaker decay signal relative to the flare peak, the 2.223 MeV signal after 100 s would still be dominated by the flare peak. 
        
%----------

        \subsubsection{Hard X-ray and $\gamma$-ray imaging results}

        Our previously introduced qualitative model shows that for the first 100 s after the start of the RHESSI observations, the 2.223 MeV signal mainly derives from protons that are accelerated at the peak of the flare (bottom panel of Fig.~\ref{fig:convolution-time-profiles}). Hence, by comparing the 2.223 MeV image reconstructed from this first part with the WL sources at the peak time, we can directly compare the precipitation site of flare-accelerated protons relative to the flare ribbons. 
        
        The left panel in Fig.~\ref{fig:rhessi-images} shows the comparison, revealing that the 2.223 MeV image produced mainly by protons from the peak time of the flare nicely agrees in location with WL. The limited dynamic range does not allow us to make a very detailed comparison; nevertheless, the strongest 2.223 MeV source is clearly associated with the easternmost part of the WL flare ribbon. The HXR images for this time interval are of course missing, but the general close association of WL and HXR flare ribbons sources reported in the literature strongly suggests that electron and proton precipitation happens on flare ribbons. Therefore, there is no evidence from RHESSI imaging that electrons and protons are precipitating at different locations. 
        
        The right panel of Fig.~\ref{fig:rhessi-images} is similar to the original figure from \citet{2006ApJ...644L..93H} comparing a time-shifted $\gamma$-ray image with a HXR image, but for a significant shorter integration time of 150 s compared to the 21.67 minutes (1300 s) considered in \citet{2006ApJ...644L..93H}. 
        Similarly to during the peak time, we can clearly see agreement between $\gamma$-ray and WL images. For the shorter integration time, the ribbon motion is slow compared to the angular resolution of the HXR and $\gamma$-ray image shown in Fig.~\ref{fig:rhessi-images}. From TRACE observations, the ribbon motion within the time interval is between 5 and 7 arcsec. Hence, the comparison between HXR and $\gamma$-ray centroids is not significantly affected by the different time evolution at these two energy ranges. Comparing the centroids at the two energy ranges give the same location, within errors. A forward fit to the visibilities gives a difference of up to 1.5 sigma ($\sim 8$ arcsec) for both eastern and western sources. Hence, within errors, HXR and $\gamma$-ray centroids are the same and associated with the WL ribbons. 

%----------

    \subsection{Comparison with other $\gamma$-ray flares}

    The event considered in this paper is the sole occurrence in the entire history of solar observations for which it is possible to reconstruct $\gamma$-ray images at distinct time intervals. Consequently, this analysis cannot be replicated for the other events documented in the literature where $\gamma$-ray imaging was possible. For these latter cases, only extended integration times are possible, and the published results, including the separation of the $\gamma$-ray and HXR centroid positions, are influenced by the different time evolutions. 

    For all the other flares (i.e., the July 23, 2002, event in \citet{2003ApJ...595L..69L}, the October 29, 2003, and November 2, 2003, events in \citet{2006ApJ...644L..93H}, and the January 20, 2005, event in \citet{2008ApJ...678L..63K}), GONG provides WL observations. We examined the WL ribbon evolution and found that, for all events, the flare ribbons intersect the centroid position of all $\gamma$-ray sources at some point during their temporal evolution, within one-sigma uncertainty of the reconstruction. This finding aligns with our previously described interpretation.

%--------------------------------------------------------------------

\section{Conclusions \label{sec:conslusions}}

In this study, we have revisited the exceptional X25 GOES class flare of October 28, 2003, observed by RHESSI, which produced the most intense $\gamma$-ray signal ever imaged of a solar flare. Despite this flare containing more total 2.223 MeV counts than all other RHESSI flares combined, only the decay phase just after the flare peak time was observed. By optimizing the integration time for image reconstruction using simulations of the neutron capture process by \citet{2003ApJ...595L..93M}, we could establish two time ranges for $\gamma$-ray imaging: one representing the flare peak time and a second during the start of the decay phase. We found a clear spatial correlation between the $\gamma$-ray sources and the HXR and WL images. We present two key conclusions: 1) the electron and proton precipitation sites coincide, at least within the RHESSI resolution, and 2) protons therefore contribute to WL formation, a factor often overlooked in previous studies. Both electrons and protons are found to be correlated with the WL and EUV flare ribbons. The source motion seen in WL and EUV between the flare peak time and the start of the decay is also seen in the $\gamma$-ray images, corroborating that protons as well as electrons deposit their energy in the flare ribbons. 

The previously reported spatial separation between electron and proton precipitation sites is a result of a difference in the relative time evolution of the $\gamma$-ray and HXR signals in combination with the motion of the sources as the flare ribbons separate. Hence, the conclusions in \citet{2003ApJ...595L..69L} and \citet{2006ApJ...644L..93H} that electrons and protons precipitate on different field lines are an overstatement. While RHESSI shows that protons precipitate into the flare ribbons, the available RHESSI observations do not have the diagnostic capabilities to investigate details of the precipitation sites along each of the flare ribbons.  By comparing the WL ribbons with the $\gamma$-ray sources of the other four events documented in the literature where $\gamma$-ray imaging was possible, we found that these findings are not exclusive to the October 28, 2003, event but applicable for these other events too. 

The agreement between $\gamma$-ray and WL sources highlights the the potential importance of protons in the WL emission mechanism. Previous observations reported a clear correlation between WL flare ribbons and the precipitation sites of electrons. Furthermore, HXR and WL flux are correlated as well with correlation coefficients up to $\sim$0.7 \citep{2016ApJ...816....6K}. Nevertheless, there is a large scatter in the correlation, with individual values differing by up to a factor of three from the fitted value. Part of the scatter could be attributed to protons that contribute to the WL formation process. 

In the $\gamma$-ray spectrum of solar flares, the neutron-capture at 2.223 MeV is the most prominent line. Therefore, it is the most attractive feature for image reconstruction, since solar flare $\gamma$-rays are orders of magnitude less intense than HXRs. Consequently, distinguishing between the contributions of the peak impulsive phase and the decay phase in this reconstruction process is of the utmost importance. This separation prevents averaging visibilities that can differ significantly. As there is currently no solar-dedicated $\gamma$-ray imager in space, future space missions should carefully consider this aspect in the design phase of the instrument.

This lack of dedicated solar $\gamma$-ray imaging capabilities in space underscores the need for new mission concepts to make progress in understanding  the importance of energetic protons in the flare energy release process. The balloon payload GRIPS \citep{2013SPIE.8862E..0WD} is the only planned imager, currently scheduled for its second flight in 2025 or 2026. Nevertheless, a dedicated space-born instrument needs to be envisioned. The Large Imaging Spectrometer for Solar Accelerated Nuclei \citep[LISSAN;][]{2023Aeros..10..985R}, as it was originally conceived -- to be part of a suite of instruments for an ESA M-class mission concept \citep[SPARK;][]{2023Aeros..10.1034R} -- is such a concept.

\begin{acknowledgements}
    The authors acknowledge Albert Shih, Brian Dennis, and Gordon Hurford for the constructive discussions and critical comments, which enhanced the quality of this paper. We also would like to thank Marina Battaglia for commenting on the final draft of the paper.
    RHESSI was a NASA small explorer mission lead and operated by UC Berkeley in close collaboration with GSFC. GONG data were acquired by instruments operated by NISP/NSO/AURA/NSF with contribution from NOAA. AFB is supported by the Swiss National Science Foundation Grant 200020\_213147. 
\end{acknowledgements}

%--------------------------------------------------------------------

% for the bibliography, at the end
\bibliographystyle{aa} % style aa.bst
\bibliography{biblio} % your references Yourfile.bib

\end{document}